\documentclass[prb,aps,floatfix,twocolumn]{revtex4-2}
\usepackage{amsmath,graphics,epsfig,color}
\usepackage{blkarray}
\usepackage{xparse,amssymb,amsfonts,tensor,bm}
\usepackage{xr-hyper}
\usepackage[bookmarks=false]{hyperref} 
\graphicspath{ {./figs/} }

\let\oldhat\hat

\renewcommand{\vec}[1]{\bm{#1}}
\renewcommand{\hat}[1]{\oldhat{\bm{#1}}}
\def\qqc{q}
\DeclareDocumentCommand{\qq}{ O{} O{} }{\vec \qqc_{#1}^{#2}}

\DeclareDocumentCommand{\lmat}{ O{} }{\hat a_{#1}} % A primitive, real space lattice vector. The index $i$ is enumerated over $d$ different vectors.
\def\imag{\textrm{i}}

\DeclareDocumentCommand{\ctrans}{ O{} }{\vec t_{#1}} % A real space lattice vector $\sum_{i=1}^d n_i \vec a_i$, where $n_i$ are the elements of $\vec n$. \\
\DeclareDocumentCommand{\qdisp}{ O{} O{} }{ \dispsym_{#1}^{#2}}
\DeclareDocumentCommand{\qdispvec}{ O{\cqvec} O{a} O{\alpha} }{ \qdisp[#1][(#2,#3)]}
\newcommand{\dispsym}{u}

\DeclareDocumentCommand{\tset}{ O{} }{\bm{\mathcal T}_{#1}}
\DeclareDocumentCommand{\ssupa}{O{}}{\hat s_{#1}} % A rank-$d$ matrix of integers, where $\hat N^{-1} \hat s$ is the sub-cell matrix.  \\
\DeclareDocumentCommand{\phin}{ O{\tset}}{\Phi_{#1}}  % Force tensor, which is the derivative of the energy with respect to  $\TT$  \\
\DeclareDocumentCommand{\supa}{ O{}}{\hat S_{#1}}
\DeclareDocumentCommand{\bqset}{ O{} O{}}{\tilde Q_{#1}^{#2}} %  A vector of all $Q_{\hat M}$ in the first Brillouin zone. \\
\DeclareDocumentCommand{\dqn}{ O{\qset} }{D_{#1}}  % Dynamical tensor, which is the derivative of the energy with respect to $\QQ$. \\
\DeclareDocumentCommand{\qset}{ O{} }{\vec Q_{#1}} % A set of $\mathcal N$ reciprocal space vectors $\vec q_{\vec m}$, where $\hat M$ is a
\newcommand{\dgg}{d\indices*{*_\Gamma^o_\Gamma^o}} 
\newcommand{\daa}{d\indices*{*_{\bar\Delta}^a_\Delta^a}} 
\newcommand{\dab}{d\indices*{*_{\bar\Delta}^a_\Delta^b}} 
\newcommand{\rdab}{d\indices*{*_{\bar\Delta}^a_\Delta^b}} 

\newcommand{\dbb}{d\indices*{*_{\bar\Delta}^b_\Delta^b}} 
\def\ndof{N}
\def\nid{N_{d}}
\def\nmeas{N_{m}}
\def\crm{\vec{C}}
\def\mdvec{\vec{v}}
\def\mdvecdim{N_{v,i}}
\def\idvec{\vec{d}}
\def\idvecdim{N_{d,i}}
\def\measv{\theta}
\def\nrepeat{N_{r}}
\begin{document}

\title {Precisely computing phonons via irreducible derivatives}
\author{Sasaank Bandi and C. A. Marianetti }
\address{ Department of Applied Physics and Applied Mathematics, Columbia University, New York, NY 10027}

\begin{abstract}

Computing phonons from first-principles is typically considered a solved
problem, yet inadequacies in existing techniques continue to yield deficient
results in systems with sensitive phonons.
Here we circumvent this issue using the lone irreducible derivative (LID) and
bundled irreducible derivative (BID) approaches to computing phonons via finite
displacements, where the former optimizes precision via energy derivatives and the latter
provides the most efficient algorithm using force derivatives.
A condition number optimized (CNO) basis for BID is derived which  guarantees the minimum 
amplification of error. 
Additionally, a hybrid LID-BID approach is formulated, where select irreducible derivatives
computed using LID replace BID results.
We illustrate our approach on two
prototypical systems with sensitive phonons: the shape memory alloy AuZn and 
metallic lithium.
Comparing our resulting phonons in the
aforementioned crystals to calculations in the literature reveals nontrivial
inaccuracies. Our approaches can be fully
automated, making them well suited for both niche systems of interest and high throughput
approaches.
%The strategy behind the irreducible approaches is to convert gains
%in efficiency into enhanced accuracy. 

\end{abstract}

\maketitle

\section{Introduction}

The Born-Oppenheimer potential characterizes the total energy of a collection
of nuclei and electrons at zero temperature, where the nuclei are localized at specific positions.
The second order Taylor series about some reference configuration provides a
vibrational Hamiltonian for the nuclei, from which the phonons may be
constructed. Given the importance of phonons for materials properties,
precisely and efficiently computing the Taylor series of the Born-Oppenheimer
potential from first-principles is critical. While perturbative approaches to computing
phonons have their merits, finite displacement approaches are agnostic to the first-principles
methodology and their implementations \cite{Fu2019014303}. Therefore, the development of 
advanced finite displacement methodologies which can knowingly balance precision and
efficiency is paramount. Previously, we introduced approaches to computing
phonons and their interactions via irreducible derivatives (ID) \cite{Fu2019014303}, 
and here we focus exclusively on phonons and introduce several important technical developments.

Our irreducible approaches begin by building a vibrational Hamiltonian purely
in terms of space group irreducible derivatives, such that all space group
symmetry and the homogeneity of space (e.g. acoustic sum rules) are satisfied by
construction. Each ID will then be associated with the smallest commensurate
supercell allowed by group theory, which will require the use of non-diagonal
supercells in general. At this stage, the IDs may be constructed via finite
difference using either the second energy derivatives or the first force
derivatives. The use of energy derivatives implies that each ID is measured
independently, and the method which isolates IDs is referred to as the lone
irreducible derivative (LID) approach. Alternatively, the use of force
derivatives enables multiple IDs to be measured simultaneously, and when the
IDs are computed in the fewest number of measurements allowed by group theory,
this is referred to as the bundled irreducible derivative (BID) approach. The
LID approach can also be applied to force derivatives, where one computes the
smallest number of IDs as possible in a given measurement, and therefore LID
with forces can be viewed as a minimal bundling approach. We refer to LID with
energy derivatives as LID$_0$ and with force derivatives as LID$_1$. Given that energy derivatives normally
converge faster than force derivatives with respect to the convergence
parameters, LID$_0$ is normally more accurate than LID$_1$ and BID for a given
set of convergence parameters.  Moreover, LID$_1$ is normally more accurate
than BID, given that bundling combines many irreducible representations. Of
course, for a given set of convergence parameters, BID is substantially more
efficient than LID$_1$ which is somewhat more efficient than LID$_0$.
Therefore, it is always preferable to use BID, but care is needed to ensure that 
proper convergence is obtained. In this paper, we derive the best possible way to 
numerically execute BID, and outline a method to selectively hybridize LID and BID
if needed.

When performing a BID calculation, a measurement basis is required to perform
finite difference calculations, and there are an infinite number of choices.
Previously, we introduced the notion of a condition number optimized (CNO)
measurement basis \cite{Fu2019014303}, in which there is a minimum amplification of error between the
measurements and the irreducible derivatives. However, we did not previously solve the
problem of how to find the measurement basis, and in this work we present the
solution for second order; demonstrating that zero amplification can be achieved.
It should be noted that mainstream
finite displacements methods to compute phonons \cite{Parlinski19974063,Alfe20092622,Togo20151} are also a type of BID
(see \cite{Fu2019014303} for a detailed discussion). However, these approaches are
not condition number optimized with respect to the irreducible derivatives, which are 
the quantities that are most directly probed by experiment (e.g. inelastic neutron scattering).

Given some first-principles theory, different methods for computing phonons may
yield different results due to sensitivity, and each method will have a
different tradeoff between efficiency and accuracy.  In the context of density functional theory (DFT),
density functional perturbation theory (DFPT) might be considered the
definitive solution, but DFPT still must be converged with respect to various
discretization parameters such as $k$-points and basis set cutoff, which may be
nontrivial.  While DFPT has been implemented using the tetrahedron method \cite{Kawamura2014094515},
where convergence was illustrated, widely available implementations
of DFPT only incorporate smearing integration methods, which are more difficult
to properly extrapolate to zero discretization given that both $k$-point density
and a smearing parameter must be varied. 
In any case, results of DFPT must be carefully scrutinized when sensitivities are present.
Finite displacement techniques utilize either the energy or the forces, which must be converged
with respect to the same discretization parameters, but each observable converges at a different
rate. An additional consideration for finite displacement techniques is convergence with respect
to the finite displacement discretization $\Delta$. If $\Delta$ is excessively small, impractical
values of DFT convergence parameters would be needed, while if $\Delta$ is excessively large, 
anharmonic contributions will deviate from leading order. Therefore, it is imperative to execute
multiple $\Delta$ and resolve the known leading order behavior (e.g. quadratic for central finite difference),
such that the derivative is precisely extrapolated to $\Delta=0$; and we refer to this process
as the construction of quadratic error tails. The substantial gains in efficiency
allowed by our irreducible derivative methods can be converted to gains in precision by properly
converging finite displacement calculations in all regards. Moreover, sensitivities are often associated
with particular phonon modes (e.g. see Section \ref{sec:results}), and our
approaches inherently isolate such sensitivities;
while conventional approaches inherently mix them, making the
practical task of converging results much more challenging.

While the LID and BID methods were developed in the context of computing
symmetrized displacement derivatives, they can be applied without modification
to arbitrary order strain derivatives; which we refer to as
$\epsilon$-LID and $\epsilon$-BID (see Ref. \cite{Mathis2022014314} for up to 4th order strain derivatives
computed using $\epsilon$-LID). 
Here we only focus on second strain derivatives (i.e. linear elastic
constants).  Given that the elastic constants dictate the linear phonon
dispersion at small $q$, precisely computing the elastic constants is an
integral component of precisely computing the phonons.  When applying $\epsilon$-LID and $\epsilon$-BID,
the strains are symmetrized according to irreducible
representations of the point group, such that the usual group theoretical
selection rules dictate the strain irreducible derivatives \emph{a priori} (i.e.
intrinsic symmetrization \cite{Fu2019014303}). For $\epsilon$-LID$_0$, the elastic
constants are computed using second energy derivatives, while $\epsilon$-LID$_1$ and 
$\epsilon$-BID use first stress derivatives; and $\epsilon$-BID measures all irreducible strain
derivatives in the fewest measurements possible via bundling.  As in the case
of phonons, $\epsilon$-LID$_0$ will normally be the most precise approach for a given set of
convergence parameters as energy derivatives are easier to converge than stress
derivatives, but $\epsilon$-BID will be the most efficient.  Conventional approaches for
performing high throughput elastic constant calculations are perhaps best considered
a type of 
$\epsilon$-LID$_1$ \cite{jong2015150009}, which is the proper philosophy for high throughput,
but $\epsilon$-LID$_0$ should be used when the definitive answer is needed.
In this study, all elastic constants are computed using $\epsilon$-LID$_0$.

To illustrate our methodological developments, we perform calculations on
the shape memory alloy AuZn and elemental lithium. 
We study AuZn in the both the cubic structure (space group
221) and the low symmetry trigonal
phase (space group 143) \cite{Makita1995430}, which is formed in a martensitic transition at
$T=64$ K \cite{Lashley2008135703}.  The phonons have been extensively explored in the cubic phase using DFT \cite{Isaeva2014104101}, though the
results had numerous sensitivities based on the details of the computational
approach.  
We also study lithium in the body centered
cubic phase, which behaves as a nearly free electron metal.  The phonons have
been measured using inelastic neutron scattering \cite{Smith1968}. Recent calculations have
found several anomalous features in the phonon spectra which are not present in
experiment \cite{Hutcheon2019014111}, and we will demonstrate that the most substantial anomalies are not present in the
numerically precise answer.

\section{BID and the CNO measurement basis}
We begin by providing the formulation of BID at second order.
Consider a real function $V(u_1,\dots,u_{\ndof})$ invariant to some ambivalent group,
where 
all independent second derivatives must be extracted. Assume that evaluating $V$ 
at some arbitrary $\{u_1,\dots,u_{\ndof}\}$ has a 
nontrivial computational cost, but also provides all first
derivatives $\{\frac{\partial V}{\partial u_i}\}$, also denoted $\{F_i\}$, subject 
to random noise.  
First order finite difference calculations may then be used to construct second derivatives, which contain
random noise from the first derivatives.
Define a ``measurement vector" as a unit vector in the $\ndof$ dimensional space of displacements.
BID is the method which 
computes
all $\nid$ independent second derivatives in the smallest number of measurements $\nmeas$.
In the
absence of symmetry, we have $\nid=\ndof(\ndof+1)/2$,
but this number will be reduced when the group
contains operations other than the identity operation.  
A measurement along the measurement vector $\measv_{i}$ will yield $\mdvecdim$
nonzero second
derivatives,
%
%$\{\frac{\partial F_j}{\partial \measv_{i} }\}$
where $\mdvecdim\le \ndof$,
which are then stacked into a vector
$\mdvec_i=(
         \frac{\partial F_1}{\partial \measv_{i} },\dots,
         \frac{\partial F_{\mdvecdim}}{\partial \measv_{i} }
         )^\intercal$.
%\begin{align*}
%\mdvec=(
%         \frac{\partial F_1}{\partial \measv_{1} },\dots,
%         \frac{\partial F_{\ndof}}{\partial \measv_{1} },
%         \dots,
%         \frac{\partial F_{\ndof}}{\partial \measv_{\nmeas} }
%         ),
%\end{align*}
A given $\mdvec_i$ can then be related to the vector
of irreducible derivatives it probes, denoted $\idvec_i$, having dimension $\idvecdim$, via the ``chain
rule matrix" $\crm_i$ as $\mdvec_i=\crm_i\idvec_i$.  The number of measurements
$\nmeas$ is chosen as the smallest number such that $\sum_{i=1}^{\nmeas} \textrm{rank}(\crm_i)\ge\nid$. 
There will be an infinite number of sets of measurement vectors, and
some criteria must be employed to select an optimum choice. 

Previously, we
introduced the notion of a condition number optimized (CNO) measurement basis \cite{Fu2019014303},
whereby the measurement vectors $\{\measv_1,\dots,\measv_{\nmeas}\}$ are
chosen to minimize the condition number of all $\crm_i$.  The condition number
of a matrix can be
computed as the ratio of the largest and smallest singular values,
and is a measure of the maximum error amplification when solving a linear
system of equations.  The BID method at second order and the CNO measurement
basis is now formally defined, but we are still left with the problem of how
to determine the CNO basis.

In order to determine the CNO basis, we consider 
the group which leaves $V(u_1,\dots,u_{\ndof})$ invariant, and we represent
this group in terms of the variables
$\{u_1,\dots,u_{\ndof}\}$; which can then be decomposed into the
irreducible representations. At second order, the Great Orthogonality Theorem dictates that
there can only be coupling between the same type of irreducible representations, and thus the key quantity will be the maximum number of instances of
a given irreducible representation,
denoted as $\nrepeat$. Given that irreducible representations of the same type must be measured separately, we have $\nmeas=\nrepeat$,
and all other irreducible representations can be partitioned into the $\nmeas$ bundles. The condition number is naturally minimized
by taking equal weights for each irreducible representation in a given bundle, allowing a condition number of one and hence
zero amplification of error. For a measurement vector with equal weights, we have
$\theta_i=\bm M_i \bm x_i$ where $\bm M_i=\frac{1}{\sqrt{N_{\theta,i}}}(1,\dots)$ and $\bm x_i$ is the
vector of all displaced irreducible representations in the bundle.
To construct the chain rule matrix, the partial derivatives $\frac{\partial x_j}{\partial \theta_i}$
are required, and these are obtained using the pseudoinverse of $\bm M_i$,
which is simply $\bm M_i^+=\bm M_i^\intercal$. The chain rule matrix
is therefore the identity matrix times $\frac{1}{\sqrt{N_{\theta,i}}}$, and the condition number is the identity.

We consider several simple examples to illustrate the preceding formulation, beginning with the 
classical coupled oscillator with mirror symmetry, having potential 
$V=\frac{1}{2}(\gamma_Ax_A^2+\gamma_Bx_B^2)$,
where $x_A$ and $x_B$ are symmetrized modes that transform like irreducible representations of the
order 2 group, and $\nid=2$. In this case, $\nrepeat=1$ and therefore $\nmeas=1$.
The condition number optimized basis is then obtained by taking equal weights of each irreducible
vector, yielding $\theta_1=\frac{1}{\sqrt2}( x_A+x_B)$. 
The irreducible derivatives $\idvec=(\gamma_A, \gamma_B)^\intercal$ 
can be extracted from the measurements 
$\mdvec=(\frac{\partial F_A}{\partial \measv_{1}},\frac{\partial F_B}{\partial \measv_{1}})^\intercal$
as $\idvec=\frac{1}{\sqrt2}\mdvec$.

The preceding example does not have repeating irreducible representations, so we now consider the three atom
oscillator with mirror symmetry, where the potential in terms of the symmetrized modes 
is 
$V=\frac{1}{2}(\gamma_Ax_A^2+\gamma_Bx_B^2+\gamma_{B'}x_{B'}^2+\gamma_{BB'}x_Bx_{B'})$ 
and $\nid=4$. In this case, $\nrepeat=2$ and therefore $\nmeas=2$. The irreducible derivative $\gamma_A$
can be bundled into either the measurement of $\gamma_B, \gamma_{BB'}$ or $\gamma_{B'}, \gamma_{BB'}$ or
both
while maintaining a condition number of one. 
Therefore, a possible condition number optimized basis is 
$\theta_1=\frac{1}{\sqrt2}( x_A+x_B)$ and $\theta_2=x_{B'}$. 
The irreducible derivatives 
$\idvec_1=(\gamma_A,\gamma_B,\gamma_{BB'})$,
are extracted from the 
measurement derivatives
$\mdvec_1=(
\frac{\partial F_A}{\partial \measv_{1}},
\frac{\partial F_B}{\partial \measv_{1}},
\frac{\partial F_{B'}}{\partial \measv_{1}}
)^\intercal$ as
$\idvec_1=\frac{1}{\sqrt2}\mdvec_1$.
The irreducible derivatives 
$\idvec_2=(\gamma_B',\gamma_{BB'})$
are extracted from the measurement derivatives
$\mdvec_2=(
\frac{\partial F_{B'}}{\partial \measv_{2}},
\frac{\partial F_B}{\partial \measv_{2}}
)^\intercal$ 
as
$\idvec_2=\mdvec_2$. While this example only contains
a one dimensional repeating irreducible representation,
repeating multidimensional irreducible representations do not require
any special attention so long as the same phase convention
is employed.

The preceding formalism and examples all pertained to ambivalent groups, which 
can always have real irreducible representations. The translation group
is not ambivalent, and will have complex irreducible representations. However,
we will demonstrate that the formulation for ambivalent groups can be applied
with trivial modifications. We consider the simplest nontrivial example, 
which can then be extended via induction. Consider the one dimensional chain
with two distinct atoms per unit cell and a system size of 3 unit cells. There will be three
$q$-points: $\Gamma$, $\Delta$, and $\bar\Delta$.  The translationally symmetrized potential
energy is given as
\begin{align}
  V=&\frac{1}{2}  \dgg u_{\Gamma}^ou_{\Gamma}^o
  +
  \daa u_{\bar\Delta}^au_{\Delta}^a+
  \dbb u_{\bar\Delta}^bu_{\Delta}^b
   \nonumber \\ & +
  (\rdab u_{\bar\Delta}^au_{\Delta}^b+ c.c.)
%  (\rdab)^* u_{\bar\Delta}^bu_{\Delta}^a
\end{align}
where $a$, $b$ label different instances of the identity representation of the little group at the 
$\Delta$ point. While $\dgg$, $\daa$, and $\dbb$ are real numbers, $\rdab$ is complex, and both 
the real and imaginary parts must be computed. Given that atoms can only be displaced on the
real axis, a change of basis is required when performing finite displacement computations. 
A unitary transformation to the real-$q$ representation is given as
\begin{align}\label{eq:realq}
\dispsym_{\qq[][c]}=\frac{1}{\sqrt 2}(\dispsym_{\qq}+\dispsym_{\bar \qq}), &&
\dispsym_{\qq[][s]}=\frac{\imag}{\sqrt 2}(\dispsym_{\bar \qq}-\dispsym_{\qq})
\end{align}
The potential can be transformed to the real-$q$ representation as
\begin{align}
\label{eq:trans_pot}
  V=&\frac{1}{2}  \dgg u_{\Gamma}^ou_{\Gamma}^o
  +
  \frac{1}{2}\daa ( 
  u_{\Delta^c}^au_{\Delta^c}^a +
  u_{\Delta^s}^au_{\Delta^s}^a 
  )\nonumber\\&
  +\frac{1}{2}\dbb ( 
  u_{\Delta^c}^bu_{\Delta^c}^b +
  u_{\Delta^s}^bu_{\Delta^s}^b 
  )\nonumber
  \\&
  +\Re(\rdab) (
  u_{\Delta^c}^au_{\Delta^c}^b+
  u_{\Delta^s}^au_{\Delta^s}^b)
  \nonumber\\&
  +\Im(\rdab) (
  u_{\Delta^c}^au_{\Delta^s}^b +
  u_{\Delta^s}^au_{\Delta^c}^b ).
\end{align}
Here we see that the number of measurements is still given by the maximum number of repeating 
irreducible representations. One choice for the condition number optimized
basis is $\theta_1=\frac{1}{\sqrt2}(u_{\Gamma}^o+u_{\Delta^c}^a)$  and $\theta_2=u_{\Delta^c}^b$.
The irreducible derivatives
$\idvec_1=(\dgg,\daa,\Re(\dab),\Im(\dab))$,
are extracted from the
measurement derivatives
$\mdvec_1=(
\frac{\partial F_\Gamma^o}{\partial \measv_{1}},
\frac{\partial F_{\Delta_c}^a}{\partial \measv_{1}},
\frac{\partial F_{\Delta_c}^b}{\partial \measv_{1}},
\frac{\partial F_{\Delta_s}^b}{\partial \measv_{1}}
)^\intercal$ as
$\idvec_1=\frac{1}{\sqrt2}\mdvec_1$.
The irreducible derivative
$\idvec_2=(\dbb)$,
is extracted from the
measurement derivatives
$\mdvec_2=(
\frac{\partial F_{\Delta_c}^b}{\partial \measv_{1}},
)^\intercal$ as
$\idvec_2=\mdvec_2$.
In summary, performing a unitary transformation to the real-$q$ basis enables the use of the procedure
outlined for ambivalent groups. 

In the preceding case, both the real and imaginary parts of the cross derivatives between irreducible representations
must be measured. An appropriate gauge transformation can make the cross derivative purely real, but this transformation
cannot be known \emph{a priori} in general. However, if the space group has a point symmetry operation that
maps $\vec q \leftrightarrow \bar{ \vec{ q}}$, the proper phase convention can be determined \emph{a priori} and then
only a real irreducible derivative must be measured. In this case, the number of measurements
will be reduced, yielding 
$ \nmeas = \lceil \nrepeat/2\rceil $. 
To illustrate this reduction, consider the potential from Eq. \ref{eq:trans_pot}  but make the
atoms equivalent, which results in a mirror plane that maps 
$\Delta \leftrightarrow \bar \Delta$. Therefore, a phase convention can be chosen
\emph{a priori} such that $\Im(\dab)=0$, and $ \nmeas =1$. A condition number optimized basis
can be chosen as $\theta_1=\frac{1}{\sqrt3}(u_{\Gamma}^o+u_{\Delta^c}^a+u_{\Delta^s}^b)$.
The irreducible derivatives
$\idvec_1=(\dgg,\daa,\dab,\dbb)$,
are extracted from the
measurement derivatives
$\mdvec_1=(
\frac{\partial F_\Gamma^o}{\partial \measv_{1}},
\frac{\partial F_{\Delta_c}^a}{\partial \measv_{1}},
\frac{\partial F_{\Delta_c}^b}{\partial \measv_{1}},
\frac{\partial F_{\Delta_s}^b}{\partial \measv_{1}}
)^\intercal$ as
$\idvec_1=\frac{1}{\sqrt3}\mdvec_1$.

The preceding examples illustrate how to generalize the process of finding the
minimal, condition number optimized measurement basis in the one dimensional
chain with two atoms per unit cell. However, there no additional considerations
for an arbitrary crystal. 

In the preceding we have outlined a condition number optimized measurement basis,
and now we present guidelines for resolving a sensitivity associated with a given
phonon mode. In our CNO basis, each irreducible derivative can be traced to
a specific bundle, and the simplest solution would be to reevaluate the choice
of $\Delta$'s and or increase the convergence parameters for that particular bundle. However, 
given that energy derivatives normally converge faster than force derivatives, one could
also recompute the problematic irreducible derivatives using LID$_0$ and use them in place
of the erroneous BID result, which we refer to as a hybrid LID-BID approach.
\begin{figure}
	\includegraphics{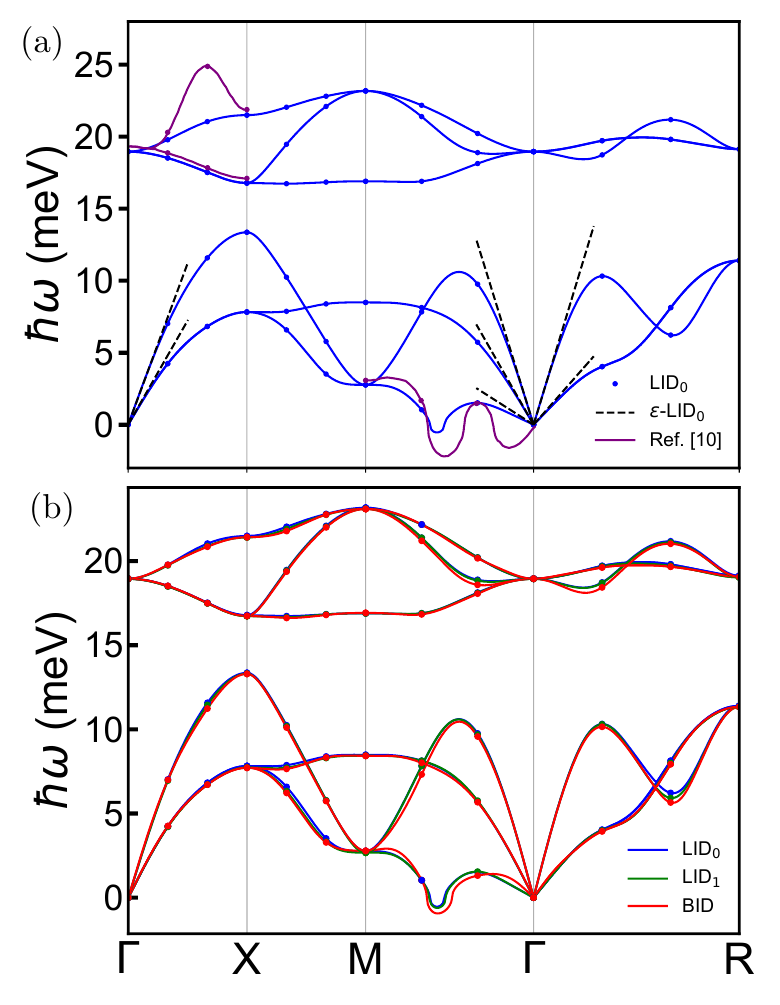}
	\caption{\label{fig:AuZn_lid} 
  DFT phonon dispersion of cubic AuZn, where points
  are computed values and lines are a Fourier interpolation. 
  $(a)$ LID$_0$ results are shown in blue, black dashed lines are the acoustic dispersion
  obtained from the elastic constants, and the results shown in purple are take from Ref. \cite{Isaeva2014104101}.
  $(b)$ LID$_0$, LID$_1$, and BID results are shown in blue, green, and red.
  }
\end{figure}

\section{Results}
\label{sec:results}
We now illustrate LID, BID, and hybrid LID-BID in several crystals where there
are discrepancies compared to the existing literature, including AuZn and body centered cubic Li.
Unless otherwise
noted, DFT calculations were performed using the Projector Augmented Wave (PAW)
method \cite{Blochl199417953,Kresse19991758}, as implemented in the Vienna
Ab-initio Simulation Package (VASP)
\cite{Kresse1993558,Kresse199414251,Kresse199615,Kresse199611169}.  
The Perdew,
Burke, Ernzerhof generalized gradient approximation (GGA) \cite{Perdew19963865}
was used for results in the main text, and   
local density approximation (LDA)\cite{Perdew19815048} results for AuZn are provided
in the Supplemental Material \cite{supmat}. 
Unless otherwise noted, a plane wave basis with a kinetic energy
cutoff of 1200 eV and 450 eV was employed for Li and AuZn, 
respectively. A $\Gamma$-centered $k$-point mesh of 
$30\times30\times30$  for both AuZn and Li.
The $k$-point integrations were done using the tetrahedron method
with Blöchl corrections\cite{Blochl199416223}.  
The DFT energies
were converged to within 10$^{-6}$ eV, while ionic relaxations were converged
to within 10$^{-5}$ eV.  
For AuZn, we used the
experimental lattice parameter of $a_0$=3.13$\textrm{\AA}$ in order to compare
with previous calculations, while energy minimization yielded a lattice parameter of 2.97 $\textrm{\AA}$ 
for Li and the relaxed trigonal structure of AuZn is provided in Supplemental Material \cite{supmat}.
For the central finite difference calculations within
LID and BID, quadratic
error tails were constructed using at least eight discretizations
(i.e. $\Delta$ in Eq. 40 in Ref. \cite{Fu2019014303}).  
It should be noted that many phonon finite difference calculations are performed with 
forward finite difference and a single discretization, and LID/BID can be executed in this manner,
though this choice would not extrapolate the discretization error to zero.
Elastic constants were measured using $\epsilon$-LID$_0$, which 
uses second strain derivatives of the energy. 
For LID/BID in Li and cubic AuZn, the Brillouin zone is discretized using a
real space supercell of
$8\times8\times8$ (i.e. multiplicity 512 and 512 atoms)
and
$6\times6\times6$ (i.e. multiplicity 216 and 432 atoms),
respectively, which are also denoted $\hat{\mathbf{S}}_{BZ}=8\hat{\mathbf{1}}$ and $\hat{\mathbf{S}}_{BZ}=6\hat{\mathbf{1}}$.
While LID/BID construct all irreducible derivatives
commensurate with $\hat{\mathbf{S}}_{BZ}=8\hat{\mathbf{1}}$ and $\hat{\mathbf{S}}_{BZ}=6\hat{\mathbf{1}}$  in Li and cubic AuZn, respectively,
all results are
extracted from supercells with multiplicity 8 and 6 \cite{Lloyd-williams2015184301,Fu2019014303}.
For LID/BID in trigonal AuZn,  the Brillouin zone is discretized 
using $\hat{\mathbf{S}}_{BZ}=2\hat{\mathbf{1}}$
(i.e. multiplicity 8 and 144 atoms), and all irreducible derivatives are extracted
using supercells with multiplicity 2.

\begin{figure}
	\includegraphics[width=\linewidth,clip=]{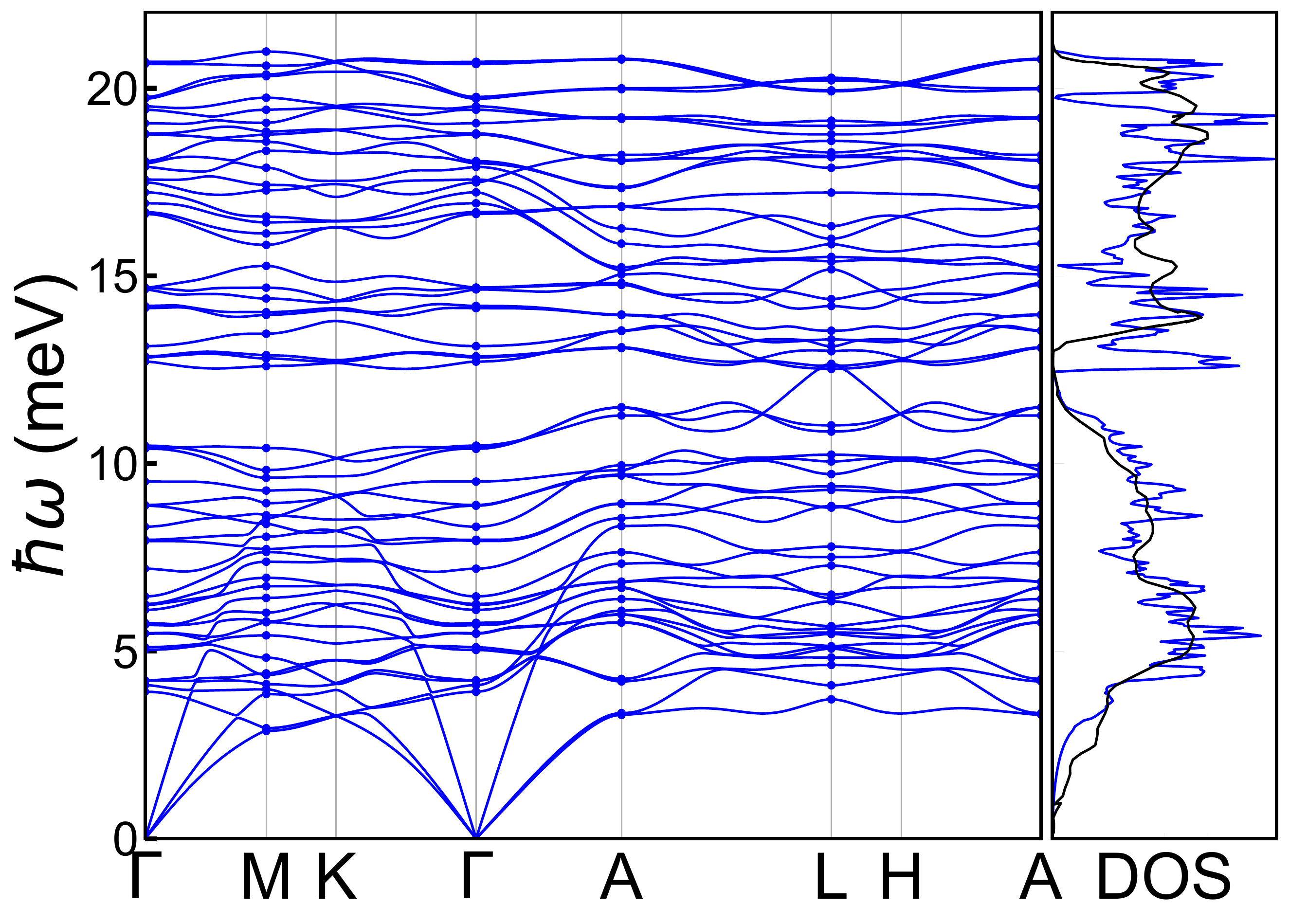}
	\caption{\label{fig:trigAuZn} ($a$) DFT phonon dispersion of trigonal AuZn computed using BID, where points
  are computed values and lines are a Fourier interpolation. ($b$)
  Phonon DOS from BID (blue) and results from Ref. \cite{Sanati2013024110} (black).
  }
\end{figure}

We begin by analyzing the cubic phase of AuZn, where experiment dictates that
there is a phase transition to the trigonal structure, which is connected to
the cubic structure via a distortion along $\vec q$ vectors in the star of
$\vec q=(\frac{1}{3},\frac{1}{3},0)$. Inelastic x-ray scattering experiments at $T=200$ K
indicate that there is a nearly soft phonon mode at $\vec q=(\frac{1}{3},\frac{1}{3},0)$ \cite{Lashley2008135703}, which
can be identified as a $B_{2}$ mode under the $C_{2v}$ little group. There are two $B_{2}$ basis modes 
at $\vec q=(\frac{1}{3},\frac{1}{3},0)$, and each
can be purely constructed of either Au or Zn, while the eigenmodes will be a linear combination.
In order to study the $B_{2}$ eigenmodes, one must compute the second derivative of each 
basis mode and the coupling between the two, resulting in three real irreducible derivatives
given that AuZn has inversion symmetry.
Therefore, when precisely computing the $B_{2}$ eigenmodes, three error tails must be carefully 
scrutinized.

%\begin{table}[htb!]
%\centering
%\begin{tabular}{|c|c|c|c|c|}
%\hline
%      & a (\AA)    & C$_{11}$ (GPa) & C$_{12}$ (GPa) & C$_{44}$ (GPa) \\ \hline
%AuZn  & 3.19       & 122.61   & 113.90   & 45.07    \\ \hline
%Li    & 3.13       & 139.91   & 136.01   & 47.37    \\ \hline
%AuZn  3.13
%Li
%\end{tabular}
%\caption{DFT relaxed lattice constants and elastic constants for cubic AuZn and body centered cubic Li computed using strain LID$_0$.}
%\end{table}

We proceed  by presenting the phonons of the cubic phase of AuZn computed
using LID$_0$ (see Figure \ref{fig:AuZn_lid}a).  
Blue points represent direct
measurements of the phonons via LID$_0$, solid blue lines are Fourier
interpolations, and dashed black lines are the linear dispersion of the
acoustic modes obtained independently from the elastic constants computed using 
$\epsilon$-LID$_0$ (see \cite{supmat} for error tails).  
Previously
published results \cite{Isaeva2014104101} using finite displacement 
are shown as purple lines for branches with
major discrepancies. Both sets of results contain all derivatives within the same
finite translation group defined by a $\hat{\mathbf{S}}_{BZ}=6\hat{\mathbf{1}}$
supercell, and therefore values from Ref. \cite{Isaeva2014104101} are measured at the same
discrete points as our calculations.  
Along the plotted directions, the
finite displacement measurements from Ref. \cite{Isaeva2014104101} are in reasonable agreement with
our own results, with the major exception of a single spurious point on the
highest optical branch between $\Gamma-X$; 
highlighting the importance of constructing error tails. 
The resulting Fourier interpolation in Ref. \cite{Isaeva2014104101} yields
imaginary frequencies for the acoustic branches near the $\Gamma$ point,
indicating that the elastic constants are negative. However,  our results clearly prove
that the elastic constants are positive, and therefore the Fourier interpolation
of the finite displacement results from Ref. \cite{Isaeva2014104101} are likely contaminated
by the spurious measurement.  
The DFPT results from Ref. \cite{Isaeva2014104101} differ
substantially from our results for the lower $B_2$ branch between
$\Gamma-M$, and the $\vec q=(\frac{1}{3},\frac{1}{3},0)$ point is predicted to be soft. The DFPT and finite displacement results must agree when both are converged.
It should be noted that a different pseudopotential was used in the DFPT calculation,
which could be the source of some differences. Additionally, the DFPT results used a smearing $k$-space integration
technique, and it is not clear if the results are converged with respect to the $k$-point density. 
While the GGA results using the VASP PAWs and the experimental lattice parameter do not yield a soft mode,
using LDA under these conditions will yield a soft mode \cite{supmat}. Additionally, using GGA with the relaxed lattice
parameter will also yield a soft mode. Therefore, the physics of the $B_{2}$ mode is somewhat sensitive and requires
a detailed investigation to provide a robust comparison with experiment.

Having established the precise phonon spectrum using LID$_0$, we now proceed to
assess the precision of BID using the CNO basis and LID$_1$ (see Figure
\ref{fig:AuZn_lid}b).  Given the importance of the $B_{2}$ modes at $\vec
q=(\frac{1}{3},\frac{1}{3},0)$, we will retain the LID$_0$ result for the two dimensional $B_2$ block, and all
other results will be obtained from BID and LID$_1$, respectively.  We see
that LID$_1$ and BID only introduce small errors, though the magnitude of the 
errors are always larger in BID, as expected.

We now proceed to the trigonal phase of AuZn, where the phonons are computed
using BID with $C_1$ symmetry (see Figure \ref{fig:trigAuZn}). The only
previous result in the literature that we are aware of is a phonon DOS
\cite{Sanati2013024110}, which we directly compare with.  There is a notable
difference in that the results from Ref. \cite{Sanati2013024110} are missing a substantial peak in the
DOS at approximately 12.75 meV. A relevant difference between our calculations
is the fact that Ref. \cite{Sanati2013024110} computes all irreducible
derivatives associated with a $2\times2\times1$ supercell while we compute all
irreducible derivatives associate with a  $2\times2\times2$ supercell. 
However, we have confirmed that there is still a peak using only the irreducible derivatives
from $2\times2\times1$ \cite{supmat}. Another major difference is that we construct
quadratic error tails while Ref. \cite{Sanati2013024110} uses a single $\Delta$,
which can cause discretization errors. A final difference is that our relaxed structure has 
some nontrivial differences from Ref. \cite{Sanati2013024110}, despite the fact that our DFT
cutoff parameters are similar, and this could account for some differences.

\begin{figure}
	\includegraphics[width=\columnwidth]{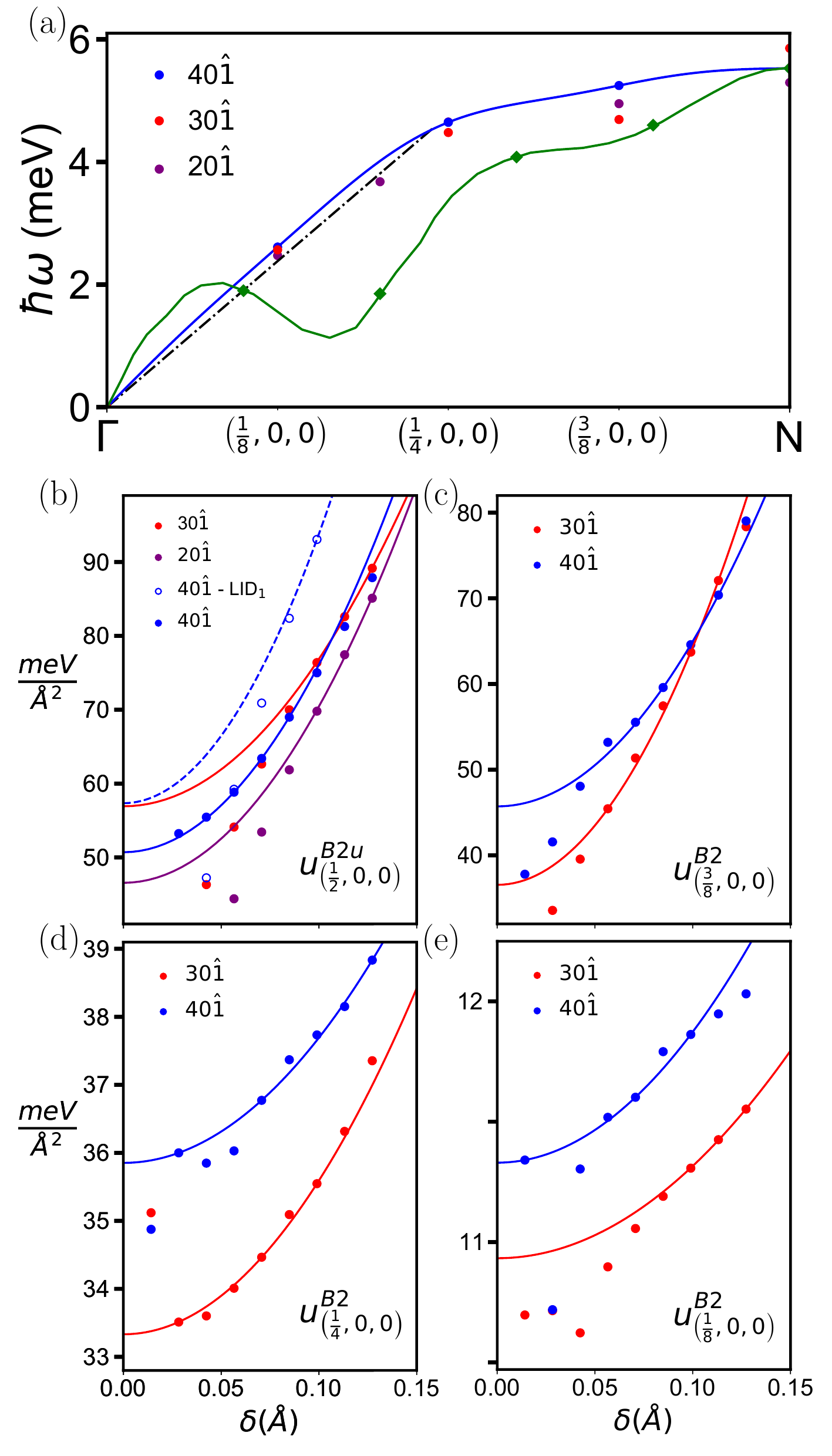}
  \caption{\label{fig:lizoom} ($a$) DFT phonons of Li for the lowest branch along $\Gamma$-$N$
  computed using BID, where points are computed values and lines are a Fourier interpolation.
  BID results are provided for $k$-point meshes of $20 \hat 1$, 
  $30 \hat 1$, and $40 \hat 1$, and results from Ref. \cite{Hutcheon2019014111} are shown in green.  
  Panels ($b$), ($c$), ($d$), and ($e$) show plots of energy vs. displacement magnitude for
  symmetrized mode amplitudes $u_{(\frac{1}{2},0,0)}^{A_{2g}}$, $u_{(\frac{3}{8},0,0)}^{B_{1u}}$,
  $u_{(\frac{1}{4},0,0)}^{B_{1u}}$, and  $u_{(\frac{1}{8},0,0)}^{B_{1u}}$,
  where $\delta$ is the real space amplitude
  which is modulated throughout the supercell as $\delta\cos(2\pi\vec q\cdot\vec t)$.
  Various $k$-point meshes are computed, and force derivatives are included in panel $b$. 
  The intercepts yield the irreducible derivative of the respective mode.}
\end{figure}
\begin{figure}
	\includegraphics[width=\columnwidth]{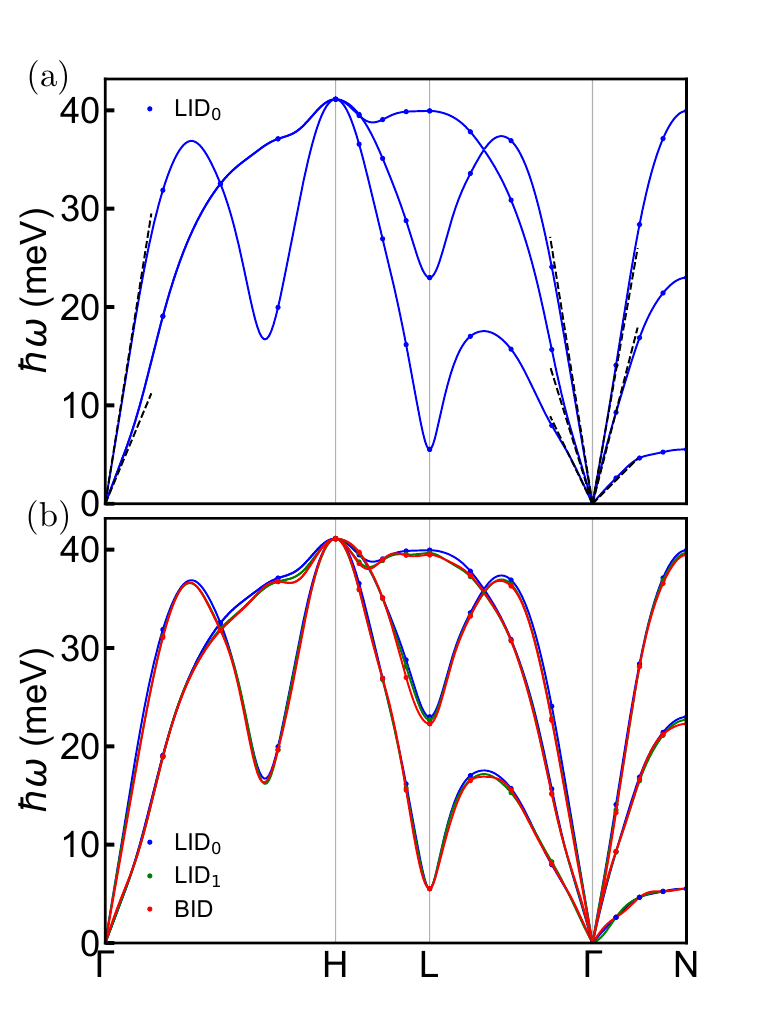}
  \caption{\label{fig:lithium} 
  DFT phonon dispersion of body centered cubic Li, where points
  are computed values and lines are a Fourier interpolation. 
  $(a)$ LID$_0$ results are shown in blue and the acoustic dispersion
  obtained from the elastic constants are black lines. 
  $(b)$ LID$_0$, LID$_1$, and BID results are shown in blue, green, and red.
  }
\end{figure}

We now consider the phonons of Li metal in the body-centered cubic
phase, and we begin by focussing on the lowest frequency
branch between $\Gamma$ and $N$ (see Figure \ref{fig:lizoom}a), as 
unusual results were obtained in previously published work \cite{Hutcheon2019014111}. 
We follow our established protocol of investigating with LID$_0$, and the task
is to ensure proper error tails are obtained, which mainly involves
including a sufficient density of $k$-points.
In Fig. \ref{fig:lizoom}$a$, circular points represent direct
measurements of the phonons via LID$_0$ at various $k$-point mesh densities, 
the blue line is a Fourier
interpolation of $\hat{\mathbf{S}}_{BZ}=8\hat{\mathbf{1}}$ using an electronic $k$-point mesh of $40\hat1$, and the dashed black line is the linear dispersion of the
acoustic mode obtained from the $\epsilon$-LID$_0$ elastic constants.  
The results of Ref. 
\cite{Hutcheon2019014111} using finite displacement calculations  with
$\hat{\mathbf{S}}_{BZ}=10\hat{\mathbf{1}}$ are shown as green diamonds and lines
for the measured and interpolated results, respectively. 
Most notably, the modulations in the Ref. \cite{Hutcheon2019014111} results
are not observed in our highest precision $40\hat1$ $k$-point mesh results.
Given that $\hat{\mathbf{S}}_{BZ}=8\hat{\mathbf{1}}$ and $\hat{\mathbf{S}}_{BZ}=10\hat{\mathbf{1}}$
are not commensurate for the interior points, we also measure $\vec q=(\frac{1}{5},0,0)$ using
a $20\hat1$ $k$-point mesh, which yields a result which still appears to be in the linear dispersion regime;
ruling out an anomaly at this particular $\vec q$ point.
The most obvious difference between our LID$_0$ calculations
and those of Ref. \cite{Hutcheon2019014111} is that the former uses second
energy derivatives and quadratic error tails while the latter uses first force derivatives 
and a single $\Delta$. Both of these differences are likely to contribute to the modulations in Ref. \cite{Hutcheon2019014111}.
However, even our results show a fair amount of variation on going from $20\hat1$ to $30\hat1$ to $40\hat1$ $k$-point mesh densities for $\vec q=(\frac{3}{8},0,0)$ and $\vec q=(\frac{1}{2},0,0)$. Therefore, it is 
interesting to explore the error tails (Fig. \ref{fig:lizoom}, panels $b$-$e$). For the case
of $\vec q =(\frac{1}{2},0,0)$ (panel $b$), the $k$-mesh of $20\hat1$ strongly deviates from a
quadratic below $\delta=0.1$ \AA, indicating that fitting larger $\delta$ to a quadratic error tail
is critical to obtaining reasonable results at this $k$-point density; and a similar assessment
holds for $30\hat1$. However, increasing to $40\hat1$ yields a clear quadratic behavior down
to $\delta=0.02$ \AA. It is also instructive to evaluate the corresponding LID$_1$ force derivative
for $40\hat1$, which does not show signs of quadratic behavior until large $\delta$, demonstrating
the limitation of force derivatives. Given the notable variations at $\vec q =(\frac{3}{8},0,0)$,
it is also useful to explore the corresponding error tails (see panel $c$), where
$30\hat1$ only has an average quadratic fit while $40\hat1$ has four consecutive
points which are strongly quadratic. The quadratic behavior for $30\hat1$ and $40\hat1$ in 
$\vec q =(\frac{1}{8},0,0)$ and $\vec q =(\frac{1}{4},0,0)$ is comparable and the overall changes
of the intercept are small (see panels $d$ and $e$). Overall, we have demonstrated that there is a notable sensitivity
to the $k$-point mesh density for $\vec q=(\frac{3}{8},0,0)$ and $\vec q=(\frac{1}{2},0,0)$.
However, $\vec q =(\frac{1}{8},0,0)$ and $\vec q =(\frac{1}{4},0,0)$ do not appear sensitive,
and the former is in reasonable agreement with the linear dispersion given by
the elastic constants.  It is also worth noting that early calculations of
phonons from DFT did not show these modulations \cite{Perdew19761421},  nor are
these features observed at $T=98$ in inelastic neutron scattering measurements
\cite{Smith1968}. 

We now proceed to compute the entire phonon spectrum using 
LID$_0$ (see Figure \ref{fig:lithium}$a$). Given the sensitivity observed in Fig. \ref{fig:lizoom},
we retain the highest precision  $40\hat1$ $k$-point mesh  results for the irreducible derivatives in this branch,
and we use $30\hat1$ otherwise. We see that the results are in good agreement with the linear
acoustic dispersions from the elastic constants at small $\vec q$.
Having established the LID$_0$ results, we now proceed
to assess the precision of LID$_1$ and BID using the CNO basis (see Figure \ref{fig:lithium}$b$). 
Once again, we retain highest precision $40\hat1$ LID$_0$ results for the irreducible derivatives
in the branch studied in Fig. \ref{fig:lizoom}.
We begin by exploring
LID$_1$ to isolated the effect of using first derivatives of the forces in
place of second derivatives of the energy.  We see that LID$_1$ only introduces
small changes, such as near the $M$-point.
These differences are small enough for our purposes, but they could be reduced
further by increasing the various convergence parameters of the calculation. 
In the BID result, where the effect of bundling can be seen by
comparing the red curve with the green curve, one can see that the magnitude
of the changes are on the same scale as the effect of using forces.  
Here we see that the large reduction in computational cost is achieved without any
appreciable loss in fidelity.

\section{Conclusion}
In summary, we have demonstrated how to accurately and efficiently compute
phonons using the LID and BID approaches.  In previous work, we defined the
notion of a condition number optimized bundled basis, but did not provide a
method to find an optimal solution \cite{Fu2019014303}.  Here we derived an
optimal solution at second order, enabling BID to provide irreducible
derivatives in the smallest number of computations with zero amplification of
error.  Typically, one will use BID to compute all irreducible derivatives. If
there are known sensitivities, or if BID error tails are deficient, LID$_0$
(i.e.  using energy derivatives) can be used to compute the problematic
irreducible derivatives, replacing the BID result. This hybrid LID-BID approach
balances accuracy and efficiency as needed. 

We demonstrated the fidelity of our irreducible approaches by addressing
sensitive phonons from the literature. In elemental Li, we computed the phonons
and confirm a sensitivity in the lowest phonon branch between $\Gamma$
and $N$, which can be properly resolved using LID$_0$ with a very fine
electronic $k$-point mesh along with quadratic error tails. In the cubic phase
of AuZn, we compute the phonons and demonstrate consistency with the elastic
constants computed using LID$_0$, and resolve the discrepancies in previous
calculations. We also compute the phonons in the trigonal phase of AuZn.  Our
irreducible methods allow one to convert gains in efficiency into gains in
accuracy, providing the definitive solution for the computed phonons in Li and
AuZn.

\section{Acknowledgements}

This work was supported by the grant DE-SC0016507 funded by the U.S. Department of Energy, Office of Science.
This research used resources of the National Energy Research Scientific Computing Center, a DOE Office of Science User Facility supported by the Office of Science of the U.S. Department of Energy under Contract No. DE-AC02-05CH11231.

\bibliography{main,supmat}

\begin{thebibliography}{25}%
\makeatletter
\providecommand \@ifxundefined [1]{%
 \@ifx{#1\undefined}
}%
\providecommand \@ifnum [1]{%
 \ifnum #1\expandafter \@firstoftwo
 \else \expandafter \@secondoftwo
 \fi
}%
\providecommand \@ifx [1]{%
 \ifx #1\expandafter \@firstoftwo
 \else \expandafter \@secondoftwo
 \fi
}%
\providecommand \natexlab [1]{#1}%
\providecommand \enquote  [1]{``#1''}%
\providecommand \bibnamefont  [1]{#1}%
\providecommand \bibfnamefont [1]{#1}%
\providecommand \citenamefont [1]{#1}%
\providecommand \href@noop [0]{\@secondoftwo}%
\providecommand \href [0]{\begingroup \@sanitize@url \@href}%
\providecommand \@href[1]{\@@startlink{#1}\@@href}%
\providecommand \@@href[1]{\endgroup#1\@@endlink}%
\providecommand \@sanitize@url [0]{\catcode `\\12\catcode `\$12\catcode
  `\&12\catcode `\#12\catcode `\^12\catcode `\_12\catcode `\%12\relax}%
\providecommand \@@startlink[1]{}%
\providecommand \@@endlink[0]{}%
\providecommand \url  [0]{\begingroup\@sanitize@url \@url }%
\providecommand \@url [1]{\endgroup\@href {#1}{\urlprefix }}%
\providecommand \urlprefix  [0]{URL }%
\providecommand \Eprint [0]{\href }%
\providecommand \doibase [0]{https://doi.org/}%
\providecommand \selectlanguage [0]{\@gobble}%
\providecommand \bibinfo  [0]{\@secondoftwo}%
\providecommand \bibfield  [0]{\@secondoftwo}%
\providecommand \translation [1]{[#1]}%
\providecommand \BibitemOpen [0]{}%
\providecommand \bibitemStop [0]{}%
\providecommand \bibitemNoStop [0]{.\EOS\space}%
\providecommand \EOS [0]{\spacefactor3000\relax}%
\providecommand \BibitemShut  [1]{\csname bibitem#1\endcsname}%
\let\auto@bib@innerbib\@empty
%</preamble>
\bibitem [{\citenamefont {Fu}\ \emph {et~al.}(2019)\citenamefont {Fu},
  \citenamefont {Kornbluth}, \citenamefont {Cheng},\ and\ \citenamefont
  {Marianetti}}]{Fu2019014303}%
  \BibitemOpen
  \bibfield  {author} {\bibinfo {author} {\bibfnamefont {L.}~\bibnamefont
  {Fu}}, \bibinfo {author} {\bibfnamefont {M.}~\bibnamefont {Kornbluth}},
  \bibinfo {author} {\bibfnamefont {Z.}~\bibnamefont {Cheng}},\ and\ \bibinfo
  {author} {\bibfnamefont {C.~A.}\ \bibnamefont {Marianetti}},\ }\bibfield
  {title} {\bibinfo {title} {Group theoretical approach to computing phonons
  and their interactions},\ }\href
  {https://doi.org/10.1103/PhysRevB.100.014303} {\bibfield  {journal} {\bibinfo
   {journal} {Phys. Rev. B}\ }\textbf {\bibinfo {volume} {100}},\ \bibinfo
  {pages} {014303} (\bibinfo {year} {2019})}\BibitemShut {NoStop}%
\bibitem [{\citenamefont {Parlinski}\ \emph {et~al.}(1997)\citenamefont
  {Parlinski}, \citenamefont {Li},\ and\ \citenamefont
  {Kawazoe}}]{Parlinski19974063}%
  \BibitemOpen
  \bibfield  {author} {\bibinfo {author} {\bibfnamefont {K.}~\bibnamefont
  {Parlinski}}, \bibinfo {author} {\bibfnamefont {Z.}~\bibnamefont {Li}},\ and\
  \bibinfo {author} {\bibfnamefont {Y.}~\bibnamefont {Kawazoe}},\ }\bibfield
  {title} {\bibinfo {title} {First-principles determination of the soft mode in
  cubic zro2},\ }\href@noop {} {\bibfield  {journal} {\bibinfo  {journal}
  {Phys. Rev. Lett.}\ }\textbf {\bibinfo {volume} {78}},\ \bibinfo {pages}
  {4063} (\bibinfo {year} {1997})}\BibitemShut {NoStop}%
\bibitem [{\citenamefont {Alfe}(2009)}]{Alfe20092622}%
  \BibitemOpen
  \bibfield  {author} {\bibinfo {author} {\bibfnamefont {D.}~\bibnamefont
  {Alfe}},\ }\bibfield  {title} {\bibinfo {title} {Phon: a program to calculate
  phonons using the small displacement method},\ }\href@noop {} {\bibfield
  {journal} {\bibinfo  {journal} {Computer Physics Communications}\ }\textbf
  {\bibinfo {volume} {180}},\ \bibinfo {pages} {2622} (\bibinfo {year}
  {2009})}\BibitemShut {NoStop}%
\bibitem [{\citenamefont {Togo}\ and\ \citenamefont
  {Tanaka}(2015)}]{Togo20151}%
  \BibitemOpen
  \bibfield  {author} {\bibinfo {author} {\bibfnamefont {A.}~\bibnamefont
  {Togo}}\ and\ \bibinfo {author} {\bibfnamefont {I.}~\bibnamefont {Tanaka}},\
  }\bibfield  {title} {\bibinfo {title} {First principles phonon calculations
  in materials science},\ }\href@noop {} {\bibfield  {journal} {\bibinfo
  {journal} {Scripta Materialia}\ }\textbf {\bibinfo {volume} {108}},\ \bibinfo
  {pages} {1} (\bibinfo {year} {2015})}\BibitemShut {NoStop}%
\bibitem [{\citenamefont {Kawamura}\ \emph {et~al.}(2014)\citenamefont
  {Kawamura}, \citenamefont {Gohda},\ and\ \citenamefont
  {Tsuneyuki}}]{Kawamura2014094515}%
  \BibitemOpen
  \bibfield  {author} {\bibinfo {author} {\bibfnamefont {M.}~\bibnamefont
  {Kawamura}}, \bibinfo {author} {\bibfnamefont {Y.}~\bibnamefont {Gohda}},\
  and\ \bibinfo {author} {\bibfnamefont {S.}~\bibnamefont {Tsuneyuki}},\
  }\bibfield  {title} {\bibinfo {title} {Improved tetrahedron method for the
  brillouin-zone integration applicable to response functions},\ }\href@noop {}
  {\bibfield  {journal} {\bibinfo  {journal} {Phys. Rev. B}\ }\textbf {\bibinfo
  {volume} {89}},\ \bibinfo {pages} {094515} (\bibinfo {year}
  {2014})}\BibitemShut {NoStop}%
\bibitem [{\citenamefont {Mathis}\ \emph {et~al.}(2022)\citenamefont {Mathis},
  \citenamefont {Khanolkar}, \citenamefont {Fu}, \citenamefont {Bryan},
  \citenamefont {Dennett}, \citenamefont {Rickert}, \citenamefont {Mann},
  \citenamefont {Winn}, \citenamefont {Abernathy}, \citenamefont {Manley},
  \citenamefont {Hurley},\ and\ \citenamefont {Marianetti}}]{Mathis2022014314}%
  \BibitemOpen
  \bibfield  {author} {\bibinfo {author} {\bibfnamefont {M.~A.}\ \bibnamefont
  {Mathis}}, \bibinfo {author} {\bibfnamefont {A.}~\bibnamefont {Khanolkar}},
  \bibinfo {author} {\bibfnamefont {L.}~\bibnamefont {Fu}}, \bibinfo {author}
  {\bibfnamefont {M.~S.}\ \bibnamefont {Bryan}}, \bibinfo {author}
  {\bibfnamefont {C.~A.}\ \bibnamefont {Dennett}}, \bibinfo {author}
  {\bibfnamefont {K.}~\bibnamefont {Rickert}}, \bibinfo {author} {\bibfnamefont
  {J.~M.}\ \bibnamefont {Mann}}, \bibinfo {author} {\bibfnamefont
  {B.}~\bibnamefont {Winn}}, \bibinfo {author} {\bibfnamefont {D.~L.}\
  \bibnamefont {Abernathy}}, \bibinfo {author} {\bibfnamefont {M.~E.}\
  \bibnamefont {Manley}}, \bibinfo {author} {\bibfnamefont {D.~H.}\
  \bibnamefont {Hurley}},\ and\ \bibinfo {author} {\bibfnamefont {C.~A.}\
  \bibnamefont {Marianetti}},\ }\bibfield  {title} {\bibinfo {title}
  {Generalized quasiharmonic approximation via space group irreducible
  derivatives},\ }\href {https://doi.org/10.1103/PhysRevB.106.014314}
  {\bibfield  {journal} {\bibinfo  {journal} {Phys. Rev. B}\ }\textbf {\bibinfo
  {volume} {106}},\ \bibinfo {pages} {014314} (\bibinfo {year}
  {2022})}\BibitemShut {NoStop}%
\bibitem [{\citenamefont {jong}\ \emph {et~al.}(2015)\citenamefont {jong},
  \citenamefont {Chen}, \citenamefont {Angsten}, \citenamefont {Jain},
  \citenamefont {Notestine}, \citenamefont {Gamst}, \citenamefont {Sluiter},
  \citenamefont {Ande}, \citenamefont {Van\_der\_zwaag}, \citenamefont {Plata},
  \citenamefont {Toher}, \citenamefont {Curtarolo}, \citenamefont {Ceder},
  \citenamefont {Persson},\ and\ \citenamefont {Asta}}]{jong2015150009}%
  \BibitemOpen
  \bibfield  {author} {\bibinfo {author} {\bibfnamefont {M.~D.}\ \bibnamefont
  {jong}}, \bibinfo {author} {\bibfnamefont {W.}~\bibnamefont {Chen}}, \bibinfo
  {author} {\bibfnamefont {T.}~\bibnamefont {Angsten}}, \bibinfo {author}
  {\bibfnamefont {A.}~\bibnamefont {Jain}}, \bibinfo {author} {\bibfnamefont
  {R.}~\bibnamefont {Notestine}}, \bibinfo {author} {\bibfnamefont
  {A.}~\bibnamefont {Gamst}}, \bibinfo {author} {\bibfnamefont
  {M.}~\bibnamefont {Sluiter}}, \bibinfo {author} {\bibfnamefont {C.~K.}\
  \bibnamefont {Ande}}, \bibinfo {author} {\bibfnamefont {S.}~\bibnamefont
  {Van\_der\_zwaag}}, \bibinfo {author} {\bibfnamefont {J.~J.}\ \bibnamefont
  {Plata}}, \bibinfo {author} {\bibfnamefont {C.}~\bibnamefont {Toher}},
  \bibinfo {author} {\bibfnamefont {S.}~\bibnamefont {Curtarolo}}, \bibinfo
  {author} {\bibfnamefont {G.}~\bibnamefont {Ceder}}, \bibinfo {author}
  {\bibfnamefont {K.~A.}\ \bibnamefont {Persson}},\ and\ \bibinfo {author}
  {\bibfnamefont {M.}~\bibnamefont {Asta}},\ }\bibfield  {title} {\bibinfo
  {title} {Charting the complete elastic properties of inorganic crystalline
  compounds},\ }\href@noop {} {\bibfield  {journal} {\bibinfo  {journal}
  {Scientific Data}\ }\textbf {\bibinfo {volume} {2}},\ \bibinfo {pages}
  {150009} (\bibinfo {year} {2015})}\BibitemShut {NoStop}%
\bibitem [{\citenamefont {Makita}\ \emph {et~al.}(1995)\citenamefont {Makita},
  \citenamefont {Nagasawa}, \citenamefont {Morii}, \citenamefont {Minakawa},\
  and\ \citenamefont {Ohno}}]{Makita1995430}%
  \BibitemOpen
  \bibfield  {author} {\bibinfo {author} {\bibfnamefont {T.}~\bibnamefont
  {Makita}}, \bibinfo {author} {\bibfnamefont {A.}~\bibnamefont {Nagasawa}},
  \bibinfo {author} {\bibfnamefont {Y.}~\bibnamefont {Morii}}, \bibinfo
  {author} {\bibfnamefont {N.}~\bibnamefont {Minakawa}},\ and\ \bibinfo
  {author} {\bibfnamefont {H.}~\bibnamefont {Ohno}},\ }\bibfield  {title}
  {\bibinfo {title} {Phonon-dispersion relations of premartensitic
  beta(1)-phase in auzn alloys},\ }\href@noop {} {\bibfield  {journal}
  {\bibinfo  {journal} {Physica B-condensed Matter}\ }\textbf {\bibinfo
  {volume} {213}},\ \bibinfo {pages} {430} (\bibinfo {year}
  {1995})}\BibitemShut {NoStop}%
\bibitem [{\citenamefont {Lashley}\ \emph {et~al.}(2008)\citenamefont
  {Lashley}, \citenamefont {Shapiro}, \citenamefont {Winn}, \citenamefont
  {Opeil}, \citenamefont {Manley}, \citenamefont {Alatas}, \citenamefont
  {Ratcliff}, \citenamefont {Park}, \citenamefont {Fisher}, \citenamefont
  {Mihaila}, \citenamefont {Riseborough}, \citenamefont {Salje},\ and\
  \citenamefont {Smith}}]{Lashley2008135703}%
  \BibitemOpen
  \bibfield  {author} {\bibinfo {author} {\bibfnamefont {J.~C.}\ \bibnamefont
  {Lashley}}, \bibinfo {author} {\bibfnamefont {S.~M.}\ \bibnamefont
  {Shapiro}}, \bibinfo {author} {\bibfnamefont {B.~L.}\ \bibnamefont {Winn}},
  \bibinfo {author} {\bibfnamefont {C.~P.}\ \bibnamefont {Opeil}}, \bibinfo
  {author} {\bibfnamefont {M.~E.}\ \bibnamefont {Manley}}, \bibinfo {author}
  {\bibfnamefont {A.}~\bibnamefont {Alatas}}, \bibinfo {author} {\bibfnamefont
  {W.}~\bibnamefont {Ratcliff}}, \bibinfo {author} {\bibfnamefont
  {T.}~\bibnamefont {Park}}, \bibinfo {author} {\bibfnamefont {R.~A.}\
  \bibnamefont {Fisher}}, \bibinfo {author} {\bibfnamefont {B.}~\bibnamefont
  {Mihaila}}, \bibinfo {author} {\bibfnamefont {P.}~\bibnamefont
  {Riseborough}}, \bibinfo {author} {\bibfnamefont {E.~K.~H.}\ \bibnamefont
  {Salje}},\ and\ \bibinfo {author} {\bibfnamefont {J.~L.}\ \bibnamefont
  {Smith}},\ }\bibfield  {title} {\bibinfo {title} {Observation of a continuous
  phase transition in a shape-memory alloy},\ }\href@noop {} {\bibfield
  {journal} {\bibinfo  {journal} {Phys. Rev. Lett.}\ }\textbf {\bibinfo
  {volume} {101}},\ \bibinfo {pages} {135703} (\bibinfo {year}
  {2008})}\BibitemShut {NoStop}%
\bibitem [{\citenamefont {Isaeva}\ \emph {et~al.}(2014)\citenamefont {Isaeva},
  \citenamefont {Souvatzis}, \citenamefont {Eriksson},\ and\ \citenamefont
  {Lashley}}]{Isaeva2014104101}%
  \BibitemOpen
  \bibfield  {author} {\bibinfo {author} {\bibfnamefont {L.}~\bibnamefont
  {Isaeva}}, \bibinfo {author} {\bibfnamefont {P.}~\bibnamefont {Souvatzis}},
  \bibinfo {author} {\bibfnamefont {O.}~\bibnamefont {Eriksson}},\ and\
  \bibinfo {author} {\bibfnamefont {J.~C.}\ \bibnamefont {Lashley}},\
  }\bibfield  {title} {\bibinfo {title} {Lattice dynamics of cubic auzn from
  first principles},\ }\href@noop {} {\bibfield  {journal} {\bibinfo  {journal}
  {Phys. Rev. B}\ }\textbf {\bibinfo {volume} {89}},\ \bibinfo {pages} {104101}
  (\bibinfo {year} {2014})}\BibitemShut {NoStop}%
\bibitem [{\citenamefont {Smith}\ \emph {et~al.}()\citenamefont {Smith},
  \citenamefont {Dolling}, \citenamefont {Nicklow}, \citenamefont
  {Vijayaraghavan},\ and\ \citenamefont {Wilkinson}}]{Smith1968}%
  \BibitemOpen
  \bibfield  {author} {\bibinfo {author} {\bibfnamefont {H.}~\bibnamefont
  {Smith}}, \bibinfo {author} {\bibfnamefont {G.}~\bibnamefont {Dolling}},
  \bibinfo {author} {\bibfnamefont {R.}~\bibnamefont {Nicklow}}, \bibinfo
  {author} {\bibfnamefont {P.}~\bibnamefont {Vijayaraghavan}},\ and\ \bibinfo
  {author} {\bibfnamefont {M.}~\bibnamefont {Wilkinson}},\ }\bibfield  {title}
  {\bibinfo {title} {Proc. conf. on inelastic neutron scattering},\ }p.\
  \bibinfo {pages} {149}\BibitemShut {NoStop}%
\bibitem [{\citenamefont {Hutcheon}\ and\ \citenamefont
  {Needs}(2019)}]{Hutcheon2019014111}%
  \BibitemOpen
  \bibfield  {author} {\bibinfo {author} {\bibfnamefont {M.}~\bibnamefont
  {Hutcheon}}\ and\ \bibinfo {author} {\bibfnamefont {R.}~\bibnamefont
  {Needs}},\ }\bibfield  {title} {\bibinfo {title} {Structural and vibrational
  properties of lithium under ambient conditions within density functional
  theory},\ }\href@noop {} {\bibfield  {journal} {\bibinfo  {journal} {Phys.
  Rev. B}\ }\textbf {\bibinfo {volume} {99}},\ \bibinfo {pages} {014111}
  (\bibinfo {year} {2019})}\BibitemShut {NoStop}%
\bibitem [{\citenamefont {Blochl}(1994)}]{Blochl199417953}%
  \BibitemOpen
  \bibfield  {author} {\bibinfo {author} {\bibfnamefont {P.~E.}\ \bibnamefont
  {Blochl}},\ }\bibfield  {title} {\bibinfo {title} {Projector augmented-wave
  method},\ }\href@noop {} {\bibfield  {journal} {\bibinfo  {journal} {Phys.
  Rev. B}\ }\textbf {\bibinfo {volume} {50}},\ \bibinfo {pages} {17953}
  (\bibinfo {year} {1994})}\BibitemShut {NoStop}%
\bibitem [{\citenamefont {Kresse}\ and\ \citenamefont
  {Joubert}(1999)}]{Kresse19991758}%
  \BibitemOpen
  \bibfield  {author} {\bibinfo {author} {\bibfnamefont {G.}~\bibnamefont
  {Kresse}}\ and\ \bibinfo {author} {\bibfnamefont {D.}~\bibnamefont
  {Joubert}},\ }\bibfield  {title} {\bibinfo {title} {From ultrasoft
  pseudopotentials to the projector augmented-wave method},\ }\href@noop {}
  {\bibfield  {journal} {\bibinfo  {journal} {Phys. Rev. B}\ }\textbf {\bibinfo
  {volume} {59}},\ \bibinfo {pages} {1758} (\bibinfo {year}
  {1999})}\BibitemShut {NoStop}%
\bibitem [{\citenamefont {Kresse}\ and\ \citenamefont
  {Hafner}(1993)}]{Kresse1993558}%
  \BibitemOpen
  \bibfield  {author} {\bibinfo {author} {\bibfnamefont {G.}~\bibnamefont
  {Kresse}}\ and\ \bibinfo {author} {\bibfnamefont {J.}~\bibnamefont
  {Hafner}},\ }\bibfield  {title} {\bibinfo {title} {Abinitio
  molecular-dynamics for liquid-metals},\ }\href@noop {} {\bibfield  {journal}
  {\bibinfo  {journal} {Phys. Rev. B}\ }\textbf {\bibinfo {volume} {47}},\
  \bibinfo {pages} {558} (\bibinfo {year} {1993})}\BibitemShut {NoStop}%
\bibitem [{\citenamefont {Kresse}\ and\ \citenamefont
  {Hafner}(1994)}]{Kresse199414251}%
  \BibitemOpen
  \bibfield  {author} {\bibinfo {author} {\bibfnamefont {G.}~\bibnamefont
  {Kresse}}\ and\ \bibinfo {author} {\bibfnamefont {J.}~\bibnamefont
  {Hafner}},\ }\bibfield  {title} {\bibinfo {title} {Ab-initio
  molecular-dynamics simulation of the liquid-metal amorphous-semiconductor
  transition in germanium},\ }\href@noop {} {\bibfield  {journal} {\bibinfo
  {journal} {Phys. Rev. B}\ }\textbf {\bibinfo {volume} {49}},\ \bibinfo
  {pages} {14251} (\bibinfo {year} {1994})}\BibitemShut {NoStop}%
\bibitem [{\citenamefont {Kresse}\ and\ \citenamefont
  {Furthmuller}(1996{\natexlab{a}})}]{Kresse199615}%
  \BibitemOpen
  \bibfield  {author} {\bibinfo {author} {\bibfnamefont {G.}~\bibnamefont
  {Kresse}}\ and\ \bibinfo {author} {\bibfnamefont {J.}~\bibnamefont
  {Furthmuller}},\ }\bibfield  {title} {\bibinfo {title} {Efficiency of
  ab-initio total energy calculations for metals and semiconductors using a
  plane-wave basis set},\ }\href@noop {} {\bibfield  {journal} {\bibinfo
  {journal} {Computational Materials Science}\ }\textbf {\bibinfo {volume}
  {6}},\ \bibinfo {pages} {15} (\bibinfo {year}
  {1996}{\natexlab{a}})}\BibitemShut {NoStop}%
\bibitem [{\citenamefont {Kresse}\ and\ \citenamefont
  {Furthmuller}(1996{\natexlab{b}})}]{Kresse199611169}%
  \BibitemOpen
  \bibfield  {author} {\bibinfo {author} {\bibfnamefont {G.}~\bibnamefont
  {Kresse}}\ and\ \bibinfo {author} {\bibfnamefont {J.}~\bibnamefont
  {Furthmuller}},\ }\bibfield  {title} {\bibinfo {title} {Efficient iterative
  schemes for ab initio total-energy calculations using a plane-wave basis
  set},\ }\href@noop {} {\bibfield  {journal} {\bibinfo  {journal} {Phys. Rev.
  B}\ }\textbf {\bibinfo {volume} {54}},\ \bibinfo {pages} {11169} (\bibinfo
  {year} {1996}{\natexlab{b}})}\BibitemShut {NoStop}%
\bibitem [{\citenamefont {Perdew}\ \emph {et~al.}(1996)\citenamefont {Perdew},
  \citenamefont {Burke},\ and\ \citenamefont {Ernzerhof}}]{Perdew19963865}%
  \BibitemOpen
  \bibfield  {author} {\bibinfo {author} {\bibfnamefont {J.~P.}\ \bibnamefont
  {Perdew}}, \bibinfo {author} {\bibfnamefont {K.}~\bibnamefont {Burke}},\ and\
  \bibinfo {author} {\bibfnamefont {M.}~\bibnamefont {Ernzerhof}},\ }\bibfield
  {title} {\bibinfo {title} {Generalized gradient approximation made simple},\
  }\href@noop {} {\bibfield  {journal} {\bibinfo  {journal} {Phys. Rev. Lett.}\
  }\textbf {\bibinfo {volume} {77}},\ \bibinfo {pages} {3865} (\bibinfo {year}
  {1996})}\BibitemShut {NoStop}%
\bibitem [{\citenamefont {Perdew}\ and\ \citenamefont
  {Zunger}(1981)}]{Perdew19815048}%
  \BibitemOpen
  \bibfield  {author} {\bibinfo {author} {\bibfnamefont {J.~P.}\ \bibnamefont
  {Perdew}}\ and\ \bibinfo {author} {\bibfnamefont {A.}~\bibnamefont
  {Zunger}},\ }\bibfield  {title} {\bibinfo {title} {Self-interaction
  correction to density-functional approximations for many-electron systems},\
  }\href {https://doi.org/10.1103/PhysRevB.23.5048} {\bibfield  {journal}
  {\bibinfo  {journal} {Phys. Rev. B}\ }\textbf {\bibinfo {volume} {23}},\
  \bibinfo {pages} {5048} (\bibinfo {year} {1981})}\BibitemShut {NoStop}%
\bibitem [{\citenamefont {Blochl}\ \emph {et~al.}(1994)\citenamefont {Blochl},
  \citenamefont {Jepsen},\ and\ \citenamefont {Andersen}}]{Blochl199416223}%
  \BibitemOpen
  \bibfield  {author} {\bibinfo {author} {\bibfnamefont {P.~E.}\ \bibnamefont
  {Blochl}}, \bibinfo {author} {\bibfnamefont {O.}~\bibnamefont {Jepsen}},\
  and\ \bibinfo {author} {\bibfnamefont {O.~K.}\ \bibnamefont {Andersen}},\
  }\bibfield  {title} {\bibinfo {title} {Improved tetrahedron method for
  brillouin-zone integrations},\ }\href
  {https://doi.org/10.1103/PhysRevB.49.16223} {\bibfield  {journal} {\bibinfo
  {journal} {Phys. Rev. B}\ }\textbf {\bibinfo {volume} {49}},\ \bibinfo
  {pages} {16223} (\bibinfo {year} {1994})}\BibitemShut {NoStop}%
\bibitem [{\citenamefont {Lloyd-williams}\ and\ \citenamefont
  {Monserrat}(2015)}]{Lloyd-williams2015184301}%
  \BibitemOpen
  \bibfield  {author} {\bibinfo {author} {\bibfnamefont {J.~H.}\ \bibnamefont
  {Lloyd-williams}}\ and\ \bibinfo {author} {\bibfnamefont {B.}~\bibnamefont
  {Monserrat}},\ }\bibfield  {title} {\bibinfo {title} {Lattice dynamics and
  electron-phonon coupling calculations using nondiagonal supercells},\
  }\href@noop {} {\bibfield  {journal} {\bibinfo  {journal} {Phys. Rev. B}\
  }\textbf {\bibinfo {volume} {92}},\ \bibinfo {pages} {184301} (\bibinfo
  {year} {2015})}\BibitemShut {NoStop}%
\bibitem [{\citenamefont {Sanati}\ \emph {et~al.}(2013)\citenamefont {Sanati},
  \citenamefont {Albers}, \citenamefont {Lookman},\ and\ \citenamefont
  {Saxena}}]{Sanati2013024110}%
  \BibitemOpen
  \bibfield  {author} {\bibinfo {author} {\bibfnamefont {M.}~\bibnamefont
  {Sanati}}, \bibinfo {author} {\bibfnamefont {R.~C.}\ \bibnamefont {Albers}},
  \bibinfo {author} {\bibfnamefont {T.}~\bibnamefont {Lookman}},\ and\ \bibinfo
  {author} {\bibfnamefont {A.}~\bibnamefont {Saxena}},\ }\bibfield  {title}
  {\bibinfo {title} {First-order versus second-order phase transformation in
  auzn},\ }\href@noop {} {\bibfield  {journal} {\bibinfo  {journal} {Physical
  Review B}\ }\textbf {\bibinfo {volume} {88}},\ \bibinfo {pages} {024110}
  (\bibinfo {year} {2013})}\BibitemShut {NoStop}%
\bibitem [{sup()}]{supmat}%
  \BibitemOpen
  \href@noop {} {}\bibinfo {note} {See Supplemental Material at [URL will be
  inserted by publisher] for error tails, plane wave cutoff and $k$-point
  convergence, select LDA results, and irreducible derivatives.}\BibitemShut
  {Stop}%
\bibitem [{\citenamefont {Perdew}\ and\ \citenamefont
  {Vosko}(1976)}]{Perdew19761421}%
  \BibitemOpen
  \bibfield  {author} {\bibinfo {author} {\bibfnamefont {J.~P.}\ \bibnamefont
  {Perdew}}\ and\ \bibinfo {author} {\bibfnamefont {S.~H.}\ \bibnamefont
  {Vosko}},\ }\bibfield  {title} {\bibinfo {title} {Phonon frequencies of
  lithium from a local effective potential},\ }\href@noop {} {\bibfield
  {journal} {\bibinfo  {journal} {Journal Of Physics F-metal Physics}\ }\textbf
  {\bibinfo {volume} {6}},\ \bibinfo {pages} {1421} (\bibinfo {year}
  {1976})}\BibitemShut {NoStop}%
\end{thebibliography}%

\end{document}